# High-performance quantum entanglement generation via cascaded second-order nonlinear processes

Zichang Zhang [1], Chenzhi Yuan [1,†], Si Shen [1], Hao Yu [1], Ruiming Zhang [1], Heqing Wang [5], Hao Li [5], You Wang [1,3], Guangwei Deng [1,4], Zhiming Wang [1,6], Lixing You [5], Zhen Wang [5], Haizhi Song [1,3], Guangcan Guo [1,4], and Qiang Zhou [1,2,4,*]

[1]*Institute of Fundamental and Frontier Sciences, University of Electronic Science and Technology of China, Chengdu 610054, China*

[2]*Yangtze Delta Region Institute (Huzhou) & School of Optoelectronic Science and Engineering, University of Electronic Science and Technology of China, Huzhou 313001, China*

[3]*Southwest Institute of Technical Physics, Chengdu 610041, China*

[4]*CAS Key Laboratory of Quantum Information, University of Science and Technology of China, Hefei 230026, China*

[5]*Shanghai Institute of Microsystem and information Technology, Chinese Academy of Sciences, Shanghai 200050, China*

[6]*Shenzhen Institute for Quantum Science and Engineering, Southern University of Science and Technology, Shenzhen 518055, China*

† c.z.yuan@uestc.edu.cn; * zhouqiang@uestc.edu.cn

**Abstract**

**In this paper, we demonstrate the generation of high-performance entangled photon-pairs in different degrees of freedom from a single piece of fiber pigtailed periodically poled LiNbO$_3$ (PPLN) waveguide. We utilize cascaded second-order nonlinear optical processes, i.e. second-harmonic generation (SHG) and spontaneous parametric down conversion (SPDC), to generate photon-pairs. Previously, the performance of the photon pairs is contaminated by Raman noise photons from the fiber pigtails. Here by integrating the PPLN waveguide with noise rejecting filters, we obtain a coincidence-to-accidental ratio (CAR) higher than 52,600 with photon-pair generation and detection rate of 52.3 kHz and 3.5 kHz, respectively. Energy-time, frequency-bin and time-bin entanglement is prepared by coherently superposing correlated two-photon states in these degrees of freedom, respectively. The energy-time entangled two-photon states achieve the maximum value of CHSH-Bell inequality of $S$=2.708±0.024 with a two-photon interference visibility of 95.74±0.86%. The frequency-bin entangled two-photon states achieve fidelity of 97.56±1.79% with a spatial quantum beating visibility of 96.85±2.46%. The time-bin entangled two-photon states achieve the maximum value of CHSH-Bell inequality of $S$=2.595±0.037 and quantum tomographic fidelity of 89.07±4.35%. Our results provide a potential candidate for quantum light source in quantum photonics.**

**Introduction**

Quantum correlated/entangled photon-pairs are serving as essential resources in quantum photonics, such as quantum key distribution (QKD)[1-4], quantum teleportation[5-7], quantum enhanced metrology[8,9], and linear optical quantum information processing (LOQC)[10-11]. Approaches, such as cascaded emissions in single-emitters[12,13] and spontaneous parametric processes[14-22], are developed to generate correlated/entangled photon-pairs. The latter one includes spontaneous parametric down-conversion (SPDC) and spontaneous four-wave mixing (SFWM), already widely utilized in several applications[23-28].

The SPDC in second-order nonlinear medium can generate bright photon-pairs, and therefore have been abundantly applied in telecom band quantum photonics. Nevertheless, the generation of photon-pairs at 1.5 $\mu$m via SPDC process is usually pumped with laser at 750 nm, thus requiring a sophisticated optical system able to manage light at quite different wavelength bands. The scheme of cascaded second-harmonic generation (SHG) and SPDC processes with two[26-28] or a single waveguide[29-31] has been developed to generate photon-pairs with all fiber-coupled deat 1.5 $\mu$m. However, the performance of such quantum light sources is limited by the spontaneous Raman scattering (SpRS) noise. Although the SpRS noise has been experimentally investegated in those sources[29], a quantum light source with low noise level is not yet demonstrated.

In this paper, we obtain high-performance entangled photon-pairs in different degrees of freedom by cascaded SHG/SPDC processes in a single piece of fiber pigtailed periodically poled LiNbO$_3$ (PPLN) waveguide. The SpRS noise photons are effectively eliminated by integrating the waveguide with noise rejecting filters. Photon-pairs with a coincidence-to-accidental ratio (CAR) higher than 52,600 are generated, with a generation rate and detection rate of 52.3 kHz and 3.5 kHz, respectively.

Entanglement in degrees of freedom of energy-time, frequency-bin and time-bin are prepared by coherently superposing correlated two-photon states. The measured visibility of the Franson interference curve is 95.74±0.86% for energy-time entanglement indicating a maximum value of CHSH-Bell inequality of $S$=2.708±0.024. The prepared frequency-bin entangled two-photon states are reconstructed with a fidelity of 97.56±1.79% by observing the spatial two-photon quantum beating with a visibility of 96.85±2.46%. The time-bin entangled two-photon states achieve a maximum value of CHSH-Bell inequality of $S$=2.595±0.037, and a fidelity of 89.07±4.35% measured via quantum state tomography. Our results show that a high performance quantum light source with cascaded second-order nonlinear processes on a photonics chip should be feasible.

**Results**

**PPLN waveguide module with noise rejecting filters.** Figure 1a shows the design of PPLN waveguide module with noise rejecting filters which consists of a bandpass filter (F1), a PPLN waveguide, a bandstop filter (F2). Two fiber pigtails are used to connect the waveguide and the noise rejecting filters. The cascaded SHG and type-0 SPDC process could take place in the PPLN waveguide with a pump light to generate the correlated/entangled photon-pairs. The noise photons generated by the pump light before the input port of the module can be removed by F1 and the ones caused by the residual pump light output from the PPLN waveguide could be effectively eliminated by F2. In our design, the fiber pigtails connecting the PPLN waveguide and noise rejecting filters are both 20-cm long - limited by our fabrication process, in which few SpRS noise photons could be generated. In principle, these SpRS noise photons could be further reduced by getting rid of fiber pigtails in a fully integrated scheme, for instance we can directly integrate F1 and F2 on the both ends of the PPLN waveguide in the future.

Table 1 gives main parameters of the PPLN waveguide module from HC Photonics. A piece of 50-mm long PPLN waveguide is fabricated by reverse proton exchange (RPE) procedure with a poled period of 19$\mu$m. The normalized conversion efficiency of SHG process is 500 %/W with pump light at 1540.46 nm. Two identical DWDMs with the central transmission wavelength at 1540.46 nm and bandwidth of ~200 GHz are used to reject the noise photons.

The single-photon level spectra of correlated photon-pairs and SpRS noise photons generated from the PPLN waveguide module are measured, as shown in Figs. 1b and 1c, respectively (see details in the spectra of noise and correlated photons in Methods section). Correlated photon-pairs can be obtained in a broadband spectrum with a full width at half-maximum (FWHM) of ~60 nm, which are contaminated with different amount of SpRS noise at different frequencies. Our results show that the ratio between photon-pairs and SpRS photons is higher than 1000 for the worst case.

**Entanglement generation with PPLN waveguide module with noise rejecting filters.** Quantum entanglement in different degrees of freedom is prepared and characterized with the setups shown in Figs. 2a-e. Figure 2a shows the experimental setup for generating correlated/entangled photon-pairs. The PPLN waveguide module is pumped by either continuous wave (CW) or pulsed laser, which is selected by an optical switch in our experiment. For both cases, the pump power is amplified, attenuated, and monitored by an erbium-doped fiber amplifier (EDFA), variable optical attenuator (VOA), and 99:1 beam splitter (BS) with a monitor power meter, respectively. A dense wavelength division multiplexer (DWDM) with the central transmission wavelength at 1540.46 nm and a passband width of ~125 GHz is employed to suppress the amplified spontaneous emission noise from the EDFA. The polarization state of pump laser is manipulated by a polarization controller (PC). A polarization beam splitter (PBS) is used to ensure the polarization alignment for maximizing the efficiency of phase matching in the PPLN waveguide. The correlated/entangled photon-pairs are generated by cascaded SHG/SPDC process. An isolator is connected to the output port of the module to reject the residual second-harmonic (SH) photons at 770 nm[32]. Figures 2b-e show setups for characterizing the quantum correlation of the generated photon-pairs, for characterizing the performance of the energy-time[33] or time-bin entanglement[34-36], for coherently manipulating two-photon state[37], for preparing and measuring the frequency-bin entanglement[37-40], respectively.

**Generation of the correlated photon-pairs.** Correlated photon-pairs are generated with setups shown in Fig. 2a, in which a CW pump at 1540.46 nm is used. The generated photon-pairs are sent into setups in Fig. 2b, in which the signal and idler photons at 1531.72 nm and 1549.34 nm are obtained by using two DWDMs with a FWHM bandwidth of ~125 GHz. The signal and idler photons are detected by two superconducting nanowire single photon detectors (SNSPDs, P-CS-6, PHOTEC, see Supplementary Materials, Section 4). A time to digital convertor (TDC, ID900, ID Quantique) is used to record the counting rates of the signal and idler photons, and the coincidence events between them.

Figure 3a shows the measured counting rate of the signal photons under different levels of pump power, which are well fitted with a quadratic polynomial curve (black line). The quadratic (red line) and linear components (blue line) are corresponding to the contributions of generated photon-pairs and noise photons, respectively. It shows that correlated photon-pairs generated in cascaded SHG/SPDC processes are dominant in the generated photons. Similar results are also obtained for idler photons (see Supplementary Materials, Fig. S1a).

Figure 3b shows the coincidence-to-accidental ratio (CAR) calculated by $CAR=C_c/A_{cc}$ under different pump power, where $C_c$ and $A_{cc}$ are the coincidence count and accidental coincidence count, respectively. The inset of Fig. 3b shows a typical measured histogram with a pump power of 2 mW, which is the accumulation of coincidence events in 20 seconds. The coincidence count is calculated by the sum of 300-ps time window covering the coincidence peak, while the accidental coincidence count is estimated by the average of three 300-ps time windows away from the coincidence peak (see Supplementary Materials, Fig. S1b). The CAR reaches a maximum value of 52,600 when pump power is 0.273 mW.

With the results of signal/idler photon counting rate and their coincidence counts, we obtain the generation rate and collection efficiency of photon-pairs via calculation (see Supplementary Materials, Fig. S1c-d). Finally, we can derive that our source achieves a photon-pair generation rate of 52.3 kHz with a CAR of 52,600, and 9.1 MHz with a CAR of 443. The calculated collection efficiencies of signal and idler photons are ~27% and ~23%, respectively, including the output efficiency of the PPLN module (~73%), and the transmission efficiencies of isolator (~89%) and the DWDMs (~67%), as well as the detection efficiency of SNSPDs (~65%).

**Performance of energy-time entanglement.** The generated photon-pairs under CW pump are sent into setups in Fig. 2c to characterize the energy-time entanglement. The signal and idler photons are first separated by DWDMs, and then pass through two unbalanced Mach-Zehnder interferometers (UMZIs, MINT, Kylia), respectively. The optical path difference between the long and short arms is 625 ps in both UMZIs, while an additional phase difference $\alpha$ or $\beta$ between the two arms can be tuned by applying a voltage on a piezo actuator. The photons from each output port of the UMZIs are detected by SNSPDs and the corresponding photon counts are recorded by the TDC.

In our experiment, three peaks appear in the coincidence measurement between signal and idler photons output from the port $A_1$ and $B_1$, respectively, in the UMZIs shown in Fig. 2c. To observe the Franson interference[33], we select the coincidence counts in the central peak with a time window of 300 ps when the phase differences $\alpha$ and $\beta$ are scanned and fixed, respectively. As shown in Fig. 3c, when $\beta$ = -0.64 rad and -1.57 rad, the coincidence counts versus $\alpha$ can be well fitted by a cosine function showing visibilities of 95.74±0.86% and 93.55±3.15%, respectively, without subtracting the accidental coincidence counts. The deviation of the visibility from unit can be attributed to the accidental coincidence, the unbalanced loss and imperfect beam splitting in the UMZIs, the instability of phases $\alpha$ and $\beta$ in the UMZIs, and the frequency instability of the pump laser. On the other hand, the signal and ilder photon counting rates are almost unchanged during the measurement, indicating that the fringes in Fig. 3c are the results of the quantum interference of energy-time entangled two-photon state. The two curves shown in Fig. 3c can be utilized to further verify the violation of CHSH-Bell inequality for the energy-time entanglement[34]. The minimum visibility of violation of the Bell inequality is 70.7%, and so the Bell inequality is violated by 29 and 7 standard deviations when $\beta$ = -0.64 rad ($S$=2.708±0.024) and -1.57 rad ($S$=2.646±0.089), respectively.

With the setups shown in Fig. 2d, we coherently manipulate the energy-time entangled two-photon state by sending it into a single UMZI (see Supplementary Materials, Section 2). The coherent manipulation can prepare a superposition state of spatial bunched and anti-bunched path-entangled states, and the complex superposition coefficients of the two states can be fully manipulated by the additional phase $\varphi$ in the UMZI. The spatial bunched path-entangled state is measured by the coincidence measurement between signal and idler photons from one output of the UMZI[37], which is shown by the red circles in Fig. 3d. We can see that such coincidence shows cosinoidal oscillation with a visibility of 94.58±0.63% when the additional phase $\varphi$ changes, indicating that the prepared state converts between the spatial bunched and anti-bunched path-entangled states. An attenuated CW laser is also injected into the UMZI shown in Fig. 2d and its single photon interference is observed, the fringe of which is shown by the rectangle curves in Fig. 3d and has a visibility of 98.10±0.01%. It is obvious that the period of the oscillation of coincidence between signal and idler photons is half that of the observed single photon interference. Such difference verifies that our coherent manipulation on energy-time entangled state is based on the interference of matter wave of the entangled two-photon state.

**Generation of frequency-bin entanglement.** As shown in Fig. 2e, the generated energy-time entangled photon-pairs are directly sent into a single UMZI, by setting the additional phase difference $\varphi=\pi$ to prepare the photon-pairs to a spatial anti-bunched path-entangled state which is exactly a frequency-bin entangled state. The frequency-bin entanglement is characterized by the spatial quantum beating[37,39,40]. The two-photon state output from the UMZI is injected into a 50:50 BS with a relative arrival time delay $\tau$ between the two paths which is controlled by a VODL (variable optical delay line). To ensure the input photons of BS are in identical polarization state, PCs and PBSs along the two optical paths are used. At the output ports of the 50/50 BS, signal and idler photons are selected by two DWDMs and are detected by SNSPDs, respectively.

Figure 4a shows the result of the spatial quantum beating. When the delay time $\tau$ is changed, the coincidence counts between signal and idler photons show modulated cosinoidal oscillation, i.e. a clear signature of the frequency-bin entanglement[37,39,40]. The experimental data in Fig. 4a can be fitted by an expression (see Supplementary Materials, Section 2)

$$C(\tau) = C_0[1 - V\operatorname{sinc}(\Omega \cdot \Delta\tau)\cos(\Delta\omega_{si} \cdot \Delta\tau + \varphi)] \tag{1}$$

where $\Delta\tau = \tau - \tau_0$ with $\tau_0$ being the intrinsic time delay between signal and idler photons; $C_0$ is a constant; $V$ is the visibility; $\operatorname{sinc}(\Omega \cdot \Delta\tau)$ describes the envelope of the spatial quantum beating and it is associated with the transmission spectra of DWDMs in Fig. 2e, approximated by a rectangular function with angular frequency bandwidth of $\Omega$; $\Delta\omega_{si}$ is the difference between the central angular frequencies of the detected signal and idler photons. The fitting gives $\Omega = 2\pi \times (116.4 \pm 2.2) \times 10^9$ rad/s and $\Delta\omega_{si} = (2.220 \pm 0.002) \times 10^{12}$ rad/s which agree well with the transmission bandwidth of 125 GHz and central wavelength of 1531.72 nm (1549.34 nm) of the DWDM for signal (idler) photons in our experiment. We also get $V$=96.85±2.46 % and phase $\varphi$=0.182±0.055 rad. According to the method in Ref. [40,41], the density matrix of the frequency-bin entangled state can be reconstructed by experimental measurements (see details in the reconstruction of frequency-bin entangled state in Methods section). Figures 4b and 4c show the real and imaginary parts of the reconstructed density matrix, which gives a target-state fidelity of 97.56±1.79% with respect to the maximally entangled state $|\psi\rangle = (|\omega_s\rangle|\omega_i\rangle + |\omega_i\rangle|\omega_s\rangle)/\sqrt{2}$.

**Generation of time-bin entanglement.** The photons entangled in time-bin can be created when periodically repeated double pulses (see preparation of double-pulsed pump light in Methods section for details) are used to pump the PPLN waveguide module in Fig. 2a. The performance of the generated time-bin entanglement also can be characterized by the experiment setup shown in Fig. 2c. In this case, all the four output ports in the two UMZIs, namely, $A_1$, $A_2$, $B_1$, and $B_2$ are used.

The UMZI can project the time-bin qubit onto the time or energy bases when the optical path difference between the long and short arms of UMZI is equal to the interval of time-bins[36]. To select different bases, a synchronous electrical signal accompanying the generation of double-pulsed pump light is introduced[34]. The Franson interference of time-bin entanglement can be observed by measuring coincidence between the projections of signal photon and idler photon on their energy bases. Figure 4d shows the three-fold coincidence counts in a time window of 300 ps between the synchronous electrical signal, and signal ($A_1$ or $A_2$) and idler ($B_1$ or $B_2$) photons when the phase $\alpha$ in one UMZI is fixed and $\beta$ in another UMZI is scanned. The coincidence counts involving the port combinations of $A_1\&B_1$, $A_1\&B_2$, $A_2\&B_1$, and $A_2\&B_2$ all show remarkable interference fringes, and the raw visibilities of them are 94.59±2.43%, 92.12±2.51%, 90.30±2.36%, 94.05±2.39%, respectively.

With those Franson interference fringes, the violation of CHSH-Bell inequality can be observed[34]. The correlation coefficient is defined as

$$E(\alpha,\beta) = \frac{\sum_{i,j}(-1)^{(i+j)}R_{A_iB_j}(\alpha,\beta)}{\sum_{i,j}R_{A_iB_j}(\alpha,\beta)} \tag{2}$$

where $R_{A_iB_j}$ is the three-fold coincidence counts involving the port combinations $A_i$ ($i$=1,2) and $B_j$ ($j$=1,2). According to the theory of Franson interference[33], $R_{A_iB_j}$ is proportional to $1+(-1)^{i+j}V\cos(\alpha+\beta)$, where $V$ is visibility of the interference fringes. The correlation coefficient calculated by Eq. (2) can be derived as $E(\alpha,\beta)=V\cos(\alpha+\beta)$ with a raw visibility of 91.75±1.30%. The maximum value of CHSH-Bell inequality can be obtained as $S=2\sqrt{2}V=2.595\pm0.037$, showing a violation of Bell inequality of up to 16 standard deviations.

The entanglement of the generated photon-pairs can be confirmed unambiguously by reconstructing the density matrix via quantum state tomography[35,36]. As shown in Fig. 2c, we choose the output ports $A_1$ and $B_1$ from the two UMZIs and calculate the density matrix by projecting the time-bin entangled two-photon states onto 16 measurement bases (see quantum state tomography of time-bin qubits in Methods section for details). The coincidence counts obtained in above projection measurements are summarized in Supplementary Materials (Section 3). As a result, we obtain the following density matrix:

$$\rho = \begin{pmatrix} 0.4527 & -0.0006-i0.0104 & 0.0367-i0.0411 & 0.3973+i0.2241 \\ -0.0006+i0.0104 & 0.0042 & -0.0015-i0.0011 & -0.0036+i0.0255 \\ 0.0367+i0.0411 & -0.0015+i0.0011 & 0.0091 & 0.0020+i0.0443 \\ 0.3973-i0.2241 & -0.0036-i0.0255 & 0.0020-i0.0443 & 0.5295 \end{pmatrix} \tag{3}$$

The real and imaginary parts of this matrix are shown graphically in Figs. 4f and 4g, respectively, from which a fidelity of 89.70±4.35% is obtained with respect to the state $|\Phi^+\rangle = (|11\rangle+|22\rangle)/\sqrt{2}$ ($|1\rangle$ and $|2\rangle$ represent the qubit in early and late time-bins, respectively). The probable main causes of this limited fidelity includes the imperfection in double-pulsed pump laser, such as nonuniform pulse intensity, instable phase difference between double pulses, and limited extinction ratio, as well as the imperfection in the UMZIs, such as the unbalanced splitting ratio in BS and instability of phase differences.

**Discussion**

In this paper, we demonstrated a high-performance entangled photon source. To show the performance of our entangled photon source, we compare the CAR and the raw detected photon-pair rate (DPPR) of our photon source with previous works in which various nonlinear optical media are employed[18,26,29,30,42-50], as shown in Fig. 5a. The comparison shows that the performance of our entangled quantum light source has orders of magnitude improvement in DPPR (CAR) than other works under the same CAR (DPPR). Recently, the periodically poled thin-film lithium niobate (TFLN) nano-waveguides achieved an remarkable performance, better than more matured, macroscopic (non-integrated) systems[44]. It could be useful for the future development of quantum photonic circuits using poled TFLN. Compared with this work, our source with PPLN waveguide module can generate more photons by two orders of magnitude under the same CAR.

Two main properties of entangled quantum light sources, integratability and high-performance on combination of CAR and DPPR, are required in quantum photonics. In this paper, we further propose a more integrated and high-performance scheme, as shown in Fig. 5b. It's a TFLN waveguide stucture with two sections connected with a high-pass intermediate waveguide with cut-off wavelength around 1 $\mu$m. The SHG and SPDC processes occur in the two sections before and after the intermediate waveguide, the cut-off property of which at 1.5 $\mu$m band can effectively remove the SpRS noise photons generated before the waveguide structure and prevent the 1.5 $\mu$m pump light from pumping SpRS process in the coupling fiber after the waveguide structure.

**Methods**

**Preparation of double-pulsed pump light.** The double-pulsed pump light used in the generation of time-bin entanglement is prepared by externally modulating the CW light which is generated from a narrow-linewidth semiconductor laser (PPCL550, PURE Photonics). We apply an arbitrary waveform generator (AWG, 70002A, Tektronix) to generate a pulsed electrical signal. In order to obtain optical pulses with high extinction ratio, the electrical signal is amplified to identical to the $V_\pi$ of the lithium niobite intensity modulator (IM, GC15MZPD7813, CETC-44) by a microwave amplifier (SHF Communication Technologies AG, SHF S126 A). A 99:1 BS combined with a photodetector is used to generate feedback signal in a controller, ensuring the IM works under an optimal bias voltage. Furthermore, for the generation of time-bin entangled photon-pairs, we utilize the AWG to generate a double-pulsed electrical signal with repetition frequency, pulse interval and single pulse width of 100 MHz, 625 ps and 125 ps, respectively.

**The spectra of noise and correlated photons.** To evaluate the SpRS noise photons and correlated photon-pairs from the PPLN waveguide module, the spectra of them were recorded by photon counting. As shown in Fig. 5c, the PPLN waveguide module is pumped by a CW laser, and a PC combined with a PBS is used to select photons with variable polarizations. A tunable filter (XTA-50/U, EXFO) is applied before a SNSPD to select photons with different frequencies for detection. The SPDC process in our PPLN waveguide satisfies type-0 phase matching, i.e., the photon-pairs created share identical and unique orientation of polarization while the polarization of the SpRS noise photons can be nearly treated as a uniform distribution over all orientations[51]. Thus, we can approximately assume that the direction of polarization with the highest and lowest photon count rates are corresponding to correlated photon-pairs accompanied with SpRS noise and only SpRS noise photons, respectively. The SpRS noise photons are filtered by adjusting PC to minimize the photon counts. As shown in Fig. 5d and 5e, the photon counts versus the pump power can be fitted well by a linear function when the central wavelength of the tunable filter was set at both 1531.72 nm and 1549.34 nm which correspond to the wavelengths of the selected signal and idler photons, respectively.

The spectra of correlated photon-pairs and SpRS noise photon shown in Figs. 1b and 1c, are obtained by scanning the central wavelength of the tunable filter when PC is adjusted to make the photon count reach the maximum and the minimum, respectively, and pump power is fixed.

**Reconstruction of frequency-bin entangled state.** Using the mothod in Ref. [40,41], we can reconstructed the density matrix of the frequency-bin entangled state from experimental measurements and the expression

$$\rho = a|\omega_s\rangle|\omega_i\rangle\langle\omega_i|\langle\omega_s| + (1-a)|\omega_i\rangle|\omega_s\rangle\langle\omega_s|\langle\omega_i| \\ + (Ve^{-i\varphi}/2)|\omega_s\rangle|\omega_i\rangle\langle\omega_s|\langle\omega_i| + (Ve^{-i\varphi}/2)|\omega_i\rangle|\omega_s\rangle\langle\omega_i|\langle\omega_s|. \quad (4)$$

It is obvious that the off-diagonal elements in Eq. (4) are determined by $V$ and $\varphi$ obtained via Eq. (1). Here, the diagonal element $a$ has a meaning of the ratio of the state $|\omega_s\rangle|\omega_i\rangle$ in the frequency-bin entangled state, and it can be estimated by measuring the counting rate of signal and idler photons from the UMZI. In our experiment, $a=0.502\pm0.001$ is obtained. The real and imaginary parts of the reconstructed density matrix are shown in Figs. 4b and 4c, respectively.

**Quantum state tomography of time-bin entanglement.** With the theory given in Ref. [36], the projection measurements of a single time-bin qubit is implemented by passing through an UMZI with the time delay difference between the two arms equal to the time interval of time-bins. By injecting a time-bin qubit with state $|\psi_0\rangle = \alpha|1\rangle + \beta|2\rangle$ ($|\alpha|^2+|\beta|^2=1$) ($|1\rangle$ and $|2\rangle$ represent the qubit in early and late time-bins, respectively) into the UMZI and detecting photons from one output port of the UMZI, the photon could be observed possibly in three time slots. Photon detection at the first (third) slot corresponds to a projection of state onto $|1\rangle$ ($|2\rangle$), namely the "time basis". Reversely, detection at the middle slot corresponds to a projection of state onto $(|1\rangle + e^{-i\theta}|2\rangle)/\sqrt{2}$, which is "energy basis" depending on the additional phase difference $\theta$ between the two arms.

In order to obtain the density matrix of the time-bin entangled photon-pairs, quantum state tomography must be implemented by 16 combinations of projection measurements between different bases ($|1\rangle, |2\rangle, |D\rangle, |R\rangle$) for for signal and idler photons, where $|D\rangle = (|1\rangle + |2\rangle)/\sqrt{2}$ and $|R\rangle = (|1\rangle + i|2\rangle)/\sqrt{2}$. Figure 5f show some typical raw data in these projection measurements. When the additional phase differences $\alpha$ and $\beta$ in the UMZIs in Fig. 2c are both set at 0, two-fold coincidence between photons from the ports A₁ and B₁ in Fig. 2c are measured, and five distinguishable peaks appear in the coincidence histogram shown in Fig. 5f-(i)[34], which correspond to the projection on single basis or the sum of projections on different bases. The three middle peaks in the Fig. 5f-(i) can further split into two or three peaks all corresponding to the projection on single basis, when three-fold coincidence is implemented between the coincidence events in these peaks and the synchronous electrical signal, as shown in Fig. 5f-(ii, iii and iv). As a consquence, we obtain the two-photon projection measurements on the following bases simultaneously: $|11\rangle, |12\rangle, |1D\rangle, |21\rangle, |22\rangle, |2D\rangle, |D1\rangle, |D2\rangle$ and $|DD\rangle$. In a similar way, we can perform the two-photon projection measurements on other bases by setting the $\alpha$ and $\beta$ at 0&π/2, π/2&0, and π/2&π/2.

**Acknowledgements**

The authors thank Professor Z. Z. Wang for helpful discussions. This work is partially supported by National Key Research and Development Program of China (Nos. 2018YFA0307400, 2019YFB2203400, 2017YFA0304000, 2018YFA0306102, 2017YFB0405100); National Natural Science Foundation of China (Nos. 61775025, 62075034, 12074058, 91836102, U19A2076, 61405030, 61704164, 62005039); Sichuan Science and Technology Program (No. 2018JY0084).

**Author details**

[1]Institute of Fundamental and Frontier Sciences, University of Electronic Science and Technology of China, Chengdu 610054, China. [2]Yangtze Delta Region Institute (Huzhou) & School of Optoelectronic Science and Engineering, University of Electronic Science and Technology of China, Huzhou 313001, China. [3]Southwest Institute of Technical Physics, Chengdu 610041, China. [4]CAS Key Laboratory of Quantum Information, University of Science and Technology of China, Hefei 230026, China. [5]Shanghai Institute of Microsystem and information Technology, Chinese Academy of Sciences, Shanghai 200050, China. [6]Shenzhen Institute for Quantum Science and Engineering, Southern University of Science and Technology, Shenzhen 518055, China

**Author contributions**

**Conflict of interest**

The authors declare that they have no conflict of interest.

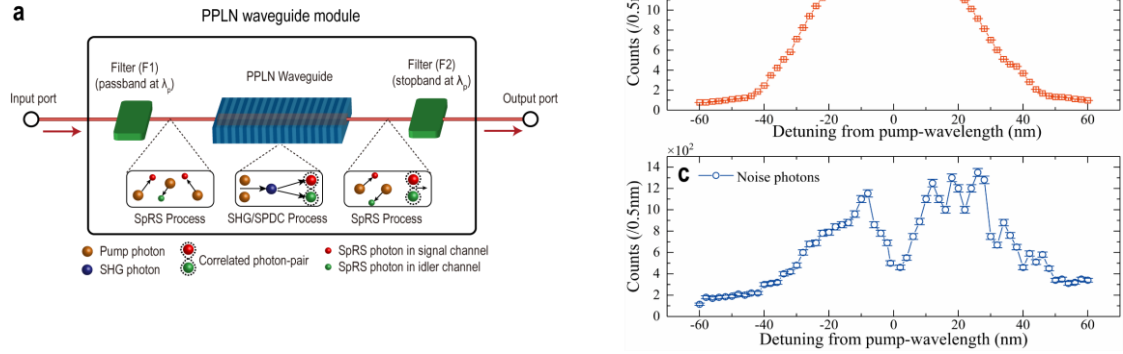

**Fig. 1 a** The design of PPLN waveguide module. The input pump photons at $\lambda_p$ pass through the filter (F1), with the input noise photons removed. The SpRS process takes place in the two fiber pigtails, while cascaded SHG/SPDC processes occur in PPLN waveguide. After filter (F2), the residual pump light is smaller than the input one by about ~20 dB, which could reduce the SpRS noise photons by ~100 times in per unit length of fiber. **b, c** Measured spectra of correlated photon-pairs and SpRS noise photons.

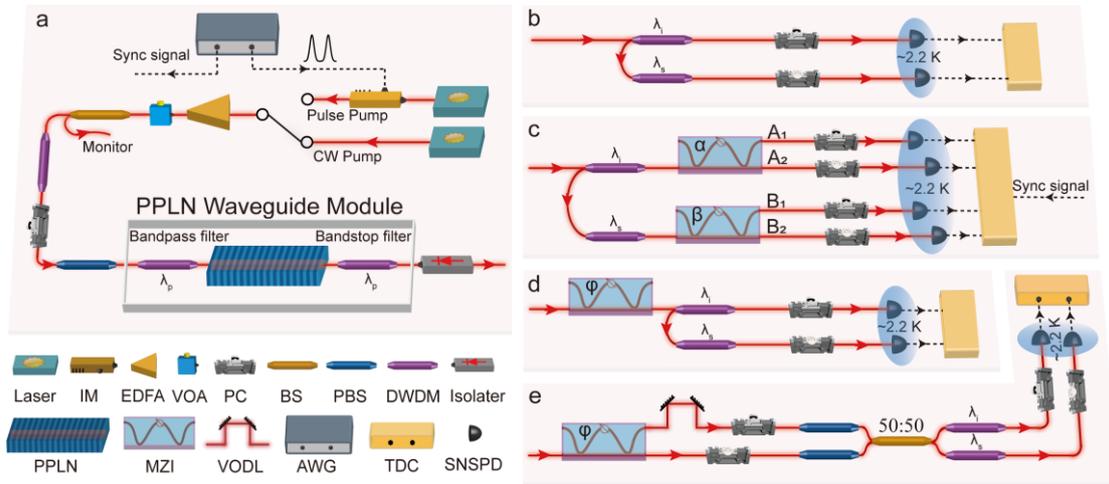

**Fig. 2 Experimental setup for the generation and measurement of correlated/entangled photon-pairs. a** Setup for preparing correlated/entangled photon-pairs. **b** Setup for characterizing correlated photon-pairs. **c** Setup for characterizing energy-time and time-bin entanglement. The synchronous signal is from AWG as shown in Fig. 2a. **d** Setup for coherently manipulating the energy-time entangled two-photon state. **e** Setup for preparing and characterizing frequency-bin entangled photon-pairs. The $\lambda_p$ in Fig. 2a is $1540.46\ nm$ and $\lambda_{s,i} = 1531.72\ nm,\ 1549.34\ nm$ in Fig. 2b-e. IM: intensity modulator; EDFA: erbium-doped fiber amplifier; VOA: variable optical attenuator; PC: polarization controller; BS: beam splitter; PBS: polarization beam splitter; DWDM: dense wavelength division multiplexer; PPLN: periodically poled LiNbO3; UMZI: unbalanced Mach-Zehnder interferometer; VODL: variable optical delay line; AWG: arbitrary waveform generator; TDC: time to digital convertor; SNSPD: superconducting nanowire single photon detector.

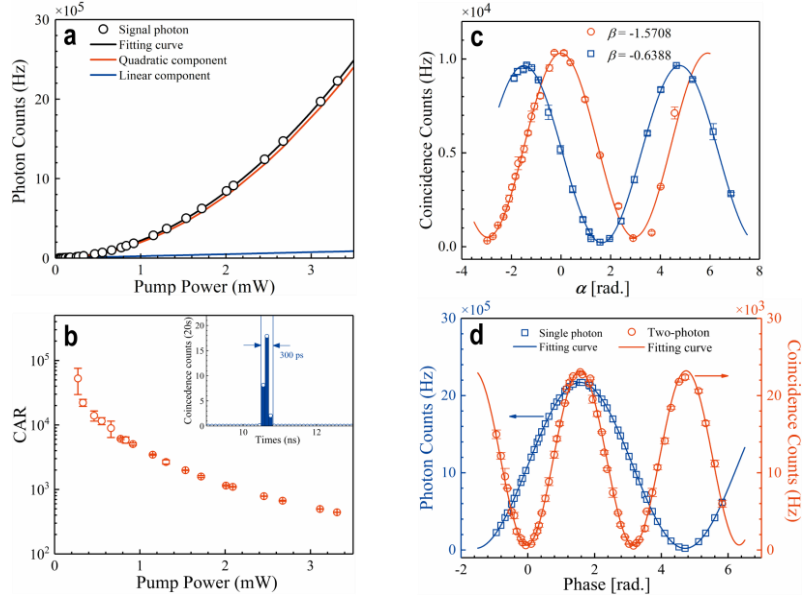

**Fig. 3 Results of correlated and energy-time entangled photon-pairs. a** Photon counting rate (black circle) in signal channel versus pump power. The black solid line is the quadratic polynomial fitting curve of the photon counting rate with the quadratic and linear parts shown as the red and blue curves, respectively. **b** Measured CAR versus pump power. The inset is the coincidence histogram between signal and idler photons when the pump power is set at 2 mW, and a coincidence window of 300 ps are marked. **c** Results of Franson interference for $\beta$ = -1.57 (red circle) and $\beta$ = -0.64 (blue rectangle). **d** Results of two-photon interference (red circle) and single photon interference (blue rectangle).

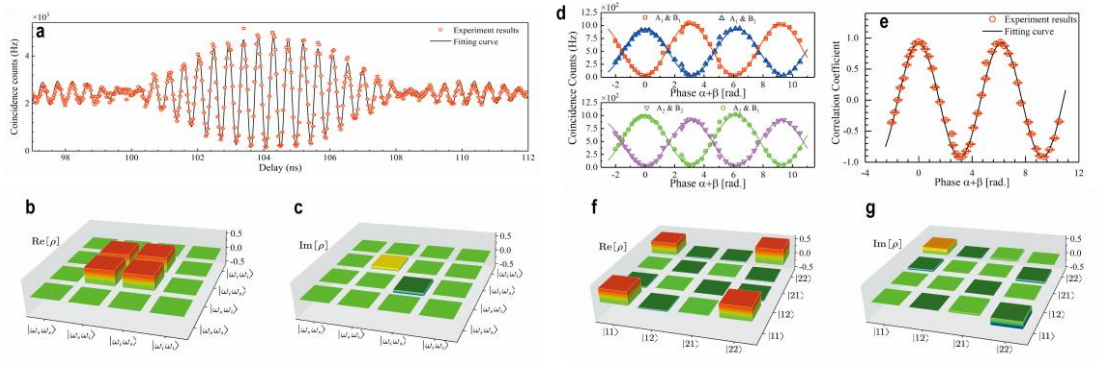

**Fig. 4 Results of frequency-bin and time-bin entangled photon-pairs. a** Spatial quantum beating of frequency-bin entangled state. **b, c** The real and the imaginary parts of the experimentally reconstructed density matrix of frequency-bin entangled photon-pairs, respectively. **d** The phase $\alpha + \beta$ dependence of the three-fold coincidence between synchronous electrical signal and the photons from the ports $A_1$&$B_1$ (red rectangle), $A_1$&$B_2$ (blue triangle), $A_2$&$B_1$ (purple inverted triangle), or $A_2$&$B_2$ (green circle). **e** Correlation coefficient $E(\alpha, \beta)$ calculated from the four curves in **d** from the four according to Eq (2). **f, g** The real and the imaginary parts of the density matrix of time-bin entangled photon-pairs, respectively.

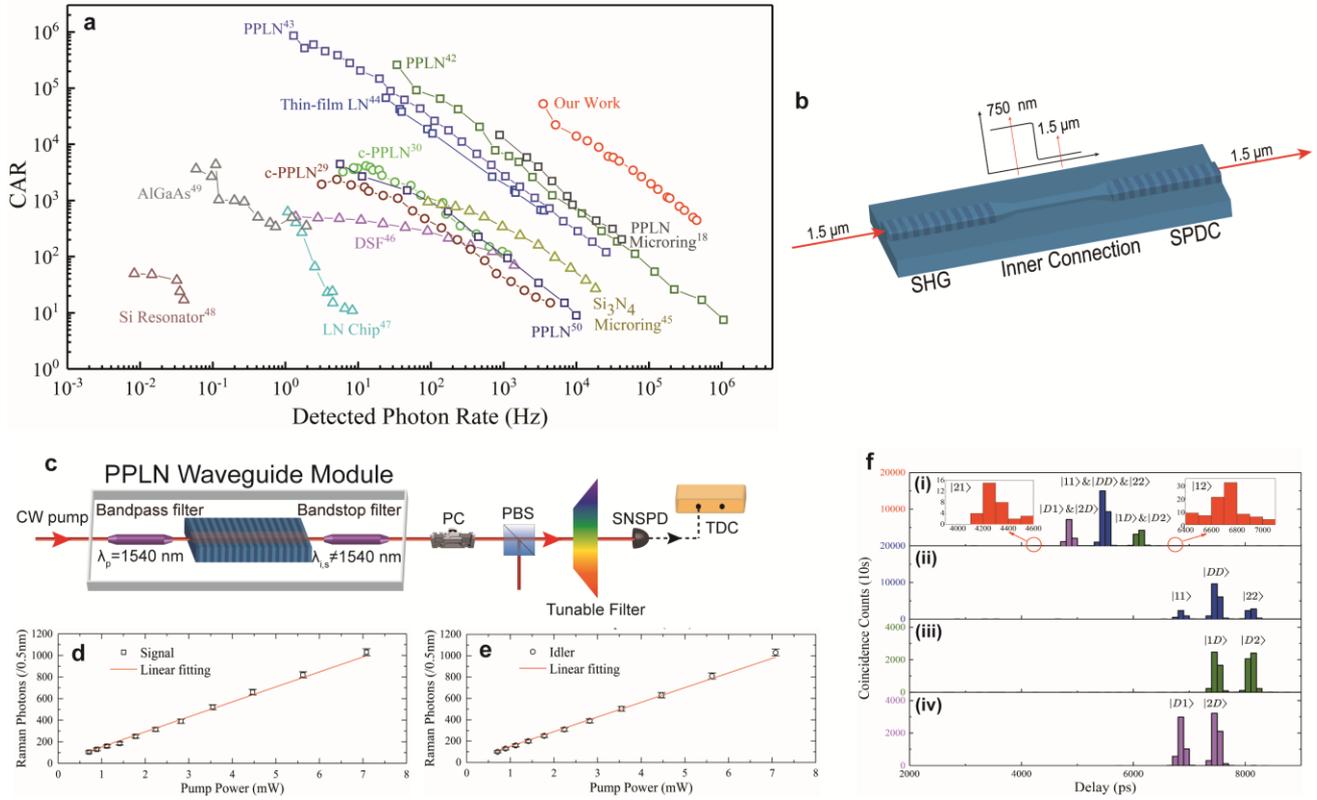

**Fig. 5 a** A comparison of the CAR values and the detected photon-pair rate among the photon-pair sources in previous literatures and our work. **b** A scheme of a fully integrated structure for generating high-performance correlated/entangled photon pairs by using cascaded SHG and SPDC processes. A inner waveguide is inserted between the SHG and SPDC regions to remove the residual pump light at 1.5 $\mu$m. **c** Setup for the spectra measurement of SpRS noise photons and correlated photon-pairs. The PBS used in this setup is a cube, rather than fiber-coupled one, in order to avoid the dependence of splitting ratio on wavelength. **d, e** The pump power dependence of the SpRS noise photons at signal and idler wavelengths, respectively. **f** A set of typical histograms of two-fold and three-fold coincidence counts obtained in the quantum state tomography of time-bin entanglement.

Table 1 Main parameters of PPLN module

| Type of Waveguide | RPE waveguide |
|---|---|
| Length of Waveguide | 50 mm |
| QPM Period | 19 $\mu$m |
| SHG normalized conversion efficiency | 500 %/W@1540.46 nm |
| Length of pigtail | 20 cm |
| Input coupling efficiency of PPLN waveguide | 73.7 % |
| Output coupling efficiency of PPLN waveguide | 85.0 % |

Supplementary information for

# High-performance quantum entanglement generation via cascaded second-order nonlinear processes

Zichang Zhang [1], Chenzhi Yuan [1,†], Si Shen [1], Hao Yu [1], Ruiming Zhang [1], Heqing Wang [5], Hao Li [5], You Wang [1,3], Guangwei Deng [1,4], Zhiming Wang [1,6], Lixing You [5], Zhen Wang [5], Haizhi Song [1,3], Guangcan Guo [1,4], and Qiang Zhou [1,2,4,*]

[1]*Institute of Fundamental and Frontier Sciences, University of Electronic Science and Technology of China, Chengdu 610054, China*

[2]*Yangtze Delta Region Institute (Huzhou) & School of Optoelectronic Science and Engineering, University of Electronic Science and Technology of China, Huzhou 313001, China*

[3]*Southwest Institute of Technical Physics, Chengdu 610041, China*

[4]*CAS Key Laboratory of Quantum Information, University of Science and Technology of China, Hefei 230026, China*

[5]*Shanghai Institute of Microsystem and information Technology, Chinese Academy of Sciences, Shanghai 200050, China*

[6]*Shenzhen Institute for Quantum Science and Engineering, Southern University of Science and Technology, Shenzhen 518055, China*

† c.z.yuan@uestc.edu.cn; * zhouqiang@uestc.edu.cn

## 1. Correlated photon-pairs

Figure S1a shows the measured counting rate of idler photons under different pump power, which are well fitted with a quadratic polynomial curve (black line). The quadratic (red line) and linear components (blue line) present the contribution of generated photon-pairs and noise photons, respectively. Figure S1b shows the measured coincidence and accidental coincidence counts in 20 seconds under different pump power levels.

The photon counts in signal channel (denoted by $N_s$) and idler channel (denoted by $N_i$), coincidence counts (denoted by $C_c$), accidental coincidence counts (denoted by $A_{cc}$) can be expressed by[1]

$$\begin{aligned}
N_s &= R\eta_s + R_s\eta_s^r + d_s, \\
N_i &= R\eta_i + R_i\eta_i^r + d_i, \\
C_c &= R\eta_s\eta_i + A_{cc}, \\
A_{cc} &= N_s N_i \Delta\tau_{bin},
\end{aligned} \tag{S1}$$

where $R$ is the generation rate of photon-pairs in PPLN waveguide; $R_s$ ($R_i$) is the SpRS photons in signal (idler) channel generated in the fiber pigtails of PPLN module; $\eta_s$ ($\eta_i$) is the collection efficiency of the signal (idler) photons; $\eta_s^r$ ($\eta_i^r$) is the collection efficiency of the SpRS photons in signal (idler) channel; $d_s$ ($d_i$) is the dark counting rates of SNSPD in signal (idler) channel; $\Delta\tau_{bin}$ is the temporal width of the coincidence window, set at 300 ps in our experiment.

Since $R$ and $R_s$ ($R_i$) Tdepend quadratically and linearly on the pump power, respectively, we can extractthe terms $R\eta_{s,i}$ and $R_{s,i}\eta_{s,i}^r$ in Eq. (S1) by selecting the quadratic and linear components in the polynomial fitting of the experimentally measured $N_{s,i}$, respectively. Then we can calculate the $R$ and $\eta_{s,i}$ under different pump power via Eq. (S1), and the results are shown in Fig. S1c and S1d.

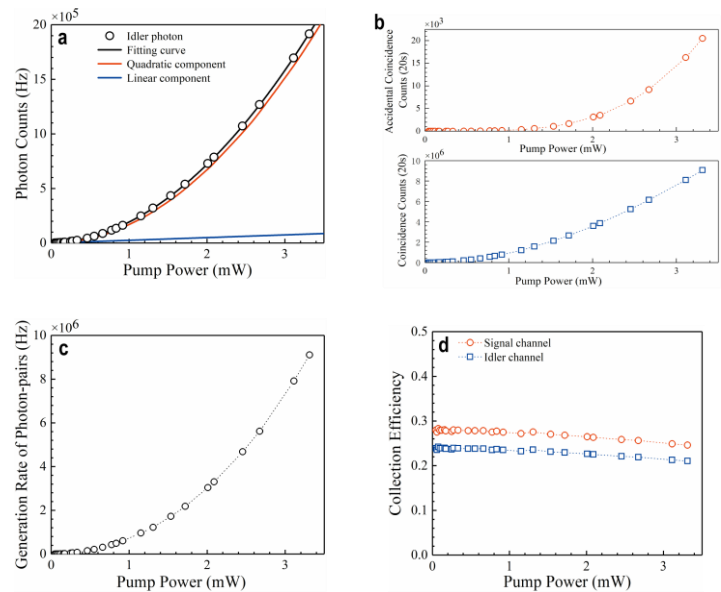

**Fig. S1 a** Photon counting rate (black circle) in idler channel versus pump power. The black solid line is the corresponding quadratic polynomial fitting curve of the photon counting rate with the quadratic and linear parts shown as the red and blue curves, respectively. **b** Measured coincidence count and accidental coincidence count versus pump power. **c** and **d** show the calculated generation rate of photon-pairs and the collection efficiencies versus pump power, respectively.

Considering the generation and transmission processes of the photon-pairs and SpRS photons, we can write $\eta_{s,i}$ as

$$\eta_{s,i} = \eta_{s,i}^c \eta_{s,i}^t \eta_{s,i}^d, \tag{S2}$$

where $\eta_{s,i}^c$ is the end-output coupling efficiency of the generated signal or idler photons from the PPLN waveguide, $\eta_{s,i}^t$ is the transmission efficiency of the filtering system for the signal and idler photons, $\eta_{s,i}^d$ is the detection efficiency of the SNSPDs. The calculated collection efficiencies of signal and idler photons are ~27% and ~23% respectively, which agree with the estimated values from Eq. (S2) with the measured $\eta_{s,i}^c$ (73%), $\eta_s^t$ (60%), $\eta_i^t$ (54%), $\eta_s^d$ (68%) and $\eta_i^d$ (63%). The slight reduction of collection efficiencies with increasing pump power could be attributed to the coincidence counting mode in our time to digital converter (TDC), which will lose some events in coincidence window when coincidence count is high.

## 2. Coherent manipulation of the energy-time entangled two-photon state

The energy-time entangled two-photon state prepared in Fig. 2a can be approximated by[2]

$$|\psi_0\rangle = |0\rangle + \sqrt{\mu_c} \int d\omega_s \int d\omega_i \varphi_0(\omega_s, \omega_i) a_s^\dagger(\omega_s) a_i^\dagger(\omega_i) |0\rangle, \tag{S3}$$

where $a_s^\dagger(\omega_s)$ and $a_i^\dagger(\omega_i)$ are the creation operators of the signal and idler photons, respectively; $\varphi_0(\omega_s, \omega_i)$ is the joint spectral amplitude of the two-photon state, defined as the product of the pump envelope function and the phase matching function[3]; $\mu_c$ is the photon flux spectral density[2] of the two-photon state. Since the first term in (S3), vacuum state $|0\rangle$, has no effect in measurement, we can only keep the second term in $|\psi_0\rangle$ in the following calculations[4].

When $|\psi_0\rangle$ is injected into the unbalanced Mach-Zehnder interferometer (UMZI) shown in Fig. S2a, the two-photon state after the first beam splitter (BS) will be[5]

$$|\psi_1\rangle = \frac{1}{\sqrt{2}}(|\psi_{11}\rangle + |\psi_{12}\rangle),$$

$$|\psi_{11}\rangle = \sqrt{\frac{\mu_c}{2}} \int d\omega_s d\omega_i \varphi_0(\omega_s, \omega_i) [a_{s,a}^\dagger(\omega_s) a_{i,a}^\dagger(\omega_i) + a_{s,b}^\dagger(\omega_s) a_{i,b}^\dagger(\omega_i)]|0\rangle, \tag{S4}$$

$$|\psi_{12}\rangle = \sqrt{\frac{\mu_c}{2}} \int d\omega_s d\omega_i \varphi_0(\omega_s, \omega_i) [a_{s,a}^\dagger(\omega_s) a_{i,b}^\dagger(\omega_i) + a_{s,b}^\dagger(\omega_s) a_{i,a}^\dagger(\omega_i)]|0\rangle,$$

where $a_{s,a}^\dagger(\omega_s)(a_{i,a}^\dagger(\omega_i))$ and $a_{s,b}^\dagger(\omega_s)(a_{i,b}^\dagger(\omega_i))$ are the creation operator of the signal (idler) photons propagating along the long arms *a* and short arm *b* in the UMZI, respectively. In the term of $|\psi_{11}\rangle$, the signal and idler photons are in the superposition state of both going along *a* or *b*, while they are in the superposition state of different paths in $|\psi_{12}\rangle$. By setting the central position of coincidence window in time domain, we can select only the state $|\psi_{11}\rangle$ via temporal filtering.

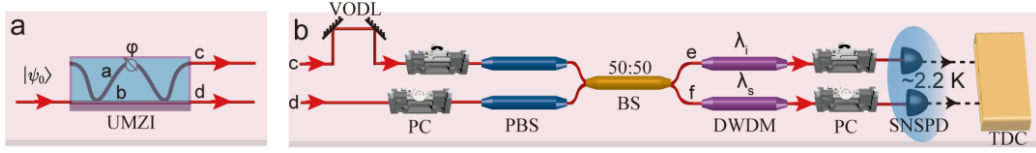

**Fig. S2 a** Structure of an UMZI. **b** Setup for characterizing frequency-bin entangled photon-pairs. UMZI: unbalanced Mach-Zehnder interferometer; VODL: variable optical delay line; PC: polarization controller; PBS: polarization beam splitter; BS: beam splitter; DWDM: dense wavelength division multiplexer; SNSPD: superconducting nanowire single photon detector; TDC: time to digital convertor.

After propagating along the two arms of the UMZI, the first term in $|\psi_1\rangle$ in Eq. (S4) evolves into

$$|\psi_2\rangle = \frac{\sqrt{\mu_c}}{2}\int d\omega_s d\omega_i \varphi_0(\omega_s,\omega_i)\left[a_{s,a}^\dagger(\omega_s)a_{i,a}^\dagger(\omega_i)e^{-i\left[(\omega_s+\omega_i)\tau_{ab}+2\varphi_a\right]} + a_{s,b}^\dagger(\omega_s)a_{i,b}^\dagger(\omega_i)\right]|0\rangle, \quad (S5)$$

where $\tau_{ab}$ is the time delay difference between the short and long arms, and $\varphi_a$ is an additional phase difference between them. In the second BS of the UMZI, the unitary transformations $a_{s,a}^\dagger(\omega_s) \to [a_{s,c}^\dagger(\omega_s)+a_{s,d}^\dagger(\omega_s)]/\sqrt{2}$ and $a_{i,b}^\dagger(\omega_i) \to [a_{i,c}^\dagger(\omega_i)-a_{i,d}^\dagger(\omega_i)]/\sqrt{2}$ are satisfied, where $a_{s,c}^\dagger(\omega_s)$ ($a_{i,c}^\dagger(\omega_s)$) and $a_{s,d}^\dagger(\omega_s)$ ($a_{i,d}^\dagger(\omega_s)$) are the creation operators of signal (idler) photon exiting from the output ports *c* and *d* of the UMZI shown in Fig. S2, respectively. After the unitary transformations, the state $|\psi_2\rangle$ in Eq. (S5) evolves into

$$|\psi_3\rangle = \frac{1}{2}\left[\left(1+e^{i2\varphi_a}\right)|\psi_b\rangle - \left(1-e^{i2\varphi_a}\right)|\psi_{ab}\rangle\right], \quad (S6)$$

where

$$|\psi_b\rangle = \frac{\sqrt{\mu_c}}{2}\int d\omega_s d\omega_i \varphi_0(\omega_s,\omega_i)e^{-i(\omega_s+\omega_i)\tau_{ab}}[a_{s,c}^\dagger(\omega_s)a_{i,c}^\dagger(\omega_i)+a_{s,d}^\dagger(\omega_s)a_{i,d}^\dagger(\omega_i)]|0\rangle, \quad (S7)$$

And

$$|\psi_{ab}\rangle = \frac{\sqrt{\mu_c}}{2}\int d\omega_s d\omega_i \varphi_0(\omega_s,\omega_i)e^{-i(\omega_s+\omega_i)\tau_{ab}}[a_{s,c}^\dagger(\omega_s)a_{i,d}^\dagger(\omega_i)+a_{s,d}^\dagger(\omega_s)a_{i,c}^\dagger(\omega_i)]|0\rangle. \quad (S8)$$

It is obvious that the states $|\psi_b\rangle$ and $|\psi_{ab}\rangle$ are spatial bunched and anti-bunched path-entangled states, respectively. According to Eq. (S6), we can switch the two-photon state between $|\psi_b\rangle$ and $|\psi_{ab}\rangle$ by setting $\varphi_b$ at $2k\pi$ and $(2k+1)\pi$ respectively.

For the preparation of energy-time entangled two-photon state, CW pump laser is used and its coherence time is long sufficiently for us to approximate the envelope function in $\varphi_0(\omega_s,\omega_i)$ by $\delta(2\omega_{p0}-\omega_s-\omega_i)$. Moreover, the phase matching function in $\varphi_0(\omega_s,\omega_i)$ can be approximated by 1 since the bandwidths of the signal and idler filters are far smaller than that of the correlated photons, as shown in Fig. 1 (b). The two approximations result in $\varphi_0(\omega_s, \omega_i) \approx \delta(2\omega_{p0} - \omega_s - \omega_i)$ and the states in Eqs. (S7) and (S8) can be written as

$$|\psi_b\rangle = \frac{\sqrt{\mu_c}}{2} e^{-i2\omega_{p0}\tau_{ab}} \int d\omega_s \left[ a_{s,c}^\dagger(\omega_s) a_{i,c}^\dagger(2\omega_{p0} - \omega_s) + a_{s,d}^\dagger(\omega_s) a_{i,d}^\dagger(2\omega_{p0} - \omega_s) \right] |0\rangle,$$

$$|\psi_{ab}\rangle = \frac{\sqrt{\mu_c}}{2} e^{-i2\omega_{p0}\tau_{ab}} \int d\omega_s \left[ a_{s,c}^\dagger(\omega_s) a_{i,d}^\dagger(2\omega_{p0} - \omega_s) + a_{s,d}^\dagger(\omega_s) a_{i,c}^\dagger(2\omega_{p0} - \omega_s) \right] |0\rangle,$$
(S9)

where $\omega_{p0}$ is the central angular frequency of the pump light in the cascaded SHG/SPDC process.

The coincidence counts between signal and idler photons from the same output port of the UMZI, for instance port c, can be expressed as

$$n_{si,c} = \eta_s \eta_i \int_T dt \int_{-\Delta\tau_{si}/2}^{\Delta\tau_{si}/2} d\tau_{si} |\langle \psi_3 | a_{s,c}^\dagger(t) a_{i,c}^\dagger(t+\tau_{si}) a_{i,c}(t+\tau_{si}) a_{s,c}(t) |\psi_3\rangle|^2,$$
(S10)

where $a_{s,c}(t) = \int d\omega_s' f_s(\omega_s') a_{s,c}(\omega_s') e^{-i\omega_s' t}/\sqrt{2\pi}$ and $a_{i,c}(t) = \int d\omega_i' f_i(\omega_i') a_{i,c}(\omega_i') e^{-i\omega_i' t}/\sqrt{2\pi}$ are the annihilation operator of the signal and idler photons from port c, with $f_s(\omega_s')$ and $f_c(\omega_s')$ being the amplitude of the transmission function of the filters in the signal and idler channels, respectively; $T$ is the integral time in coincidence measurement; $\tau_{si}$ is a time delay between the signal and idler photon detection events recorded in TDC and $\Delta\tau_{si}$ is the width of the coincidence window; $\eta_{s,i}$ summaries the collection and detection efficiencies of the signal and idler photons, respectively. In our experiment, according to the measured transmission spectra of the DWDMs used in Fig. 2d, we can express $f_{s,i}(\omega_{s,i}')$ as $f_{s,i}(\omega_{s,i}') = \text{rect}((\omega_{s,i}' - \omega_{s0,i0})/\Omega)$, which is a rectangular function with central angular frequency of $\omega_{s0,i0}$ and angular frequency bandwidth of $\Omega$. Here, $\omega_{s0,i0}$ satisfies $\omega_{s0} + \omega_{i0} = 2\omega_{p0}$.

After substituting Eqs. (S6), (S9) and the expression of $f_{s,i}(\omega_{s,i}')$ into Eq. (S10), we can get

$$n_{si,c} = \eta_s \eta_i \int_T dt \int_{-\Delta\tau_{si}/2}^{\Delta\tau_{si}/2} d\tau_{si} |\langle 0| a_{i,c}(t+\tau_{si}) a_{s,c}(t) |\psi_3\rangle|^2 = \frac{\eta_s \eta_i \mu_c}{8} [1 + \cos(2\omega_{p0}\tau_{ab} + 2\varphi_a)].$$
(S11)

In the derivation of Eq. (S11), the condition $\Delta\tau_{si} \gg 2\pi/\Omega$ is used and the integral range about $\tau_{si}$ is approximated by minus infinite to positive infinite. This condition is satisfied in our experiment with $\Delta\tau_{si} = 300\ ps$ and $2\pi/\Omega \approx 8\ ps$. In the similar way, we can get the expression of the coincidence counts between the signal and idler photons from ports c and d of the UMZI, respectively, as

$$n_{si,cd} = \frac{\eta_s \eta_i \mu_c}{8} [1 - \cos(2\omega_{p0}\tau_{ab} + 2\varphi_a)].$$
(S12)

The Eq. (S11) can describe the oscillation of coincidence counts shown by the red circles in Fig. 3d, except that the visibility in this figure is lower than 1 because of the noise photons, and the imperfect BSs and phase instability in UMZI. Also, it is obvious that $n_{si,c}$ ($n_{si,cd}$) is proportional to the projection probability of $|\psi_3\rangle$ on the component of $\int d\omega_s a_{s,c}^\dagger(\omega_s) a_{i,c}^\dagger(2\omega_{p0}-\omega_s)|0\rangle$ ($\int d\omega_s a_{s,c}^\dagger(\omega_s) a_{i,d}^\dagger(2\omega_{p0}-\omega_i)|0\rangle$) in $|\psi_b\rangle$ ($|\psi_{ab}\rangle$) shown in Eq. (S9), and therefore the oscillation of coincidence counts in Fig. 3d actually results from the oscillation of the coherently manipulated two photon state between the spatial bunched and anti-bunched path-entangled states. According to the Eq. (S9), $|\psi_{ab}\rangle$, the spatial anti-bunched path-entangled state is a frequency entangled state, a superposition of two states in which the paths c and d are occupied by photons with angular frequency of $\omega_s$ and $2\omega_{p0}-\omega_s$, respectively, or verse vice. Through the transmission band of the DWDMs used for signal and idler photons filtering, only the signal (idler) photons falling into a frequency bin centering at $\omega_{s0}$ ($2\omega_{p0}-\omega_{s0}$) are detected. This indicates that a frequency-bin entangled state is obtained by post-selection on $|\psi_{ab}\rangle$.

Figure S2b shows the setup of spatial quantum beating for characterizing the frequency-bin entangled photon-pairs. After the frequency entangled state $|\psi_{ab}\rangle$ is prepared, it is injected into a 50:50 BS by connecting the ports c and d of the UMZI as shown in Fig. S2a to the two input ports in Fig. S2b, respectively. A relative time delay $\tau$ is introduced between the paths c and d, and before the BS $|\psi_{ab}\rangle$ evolves into

$$|\psi_4\rangle = \frac{\sqrt{\mu_c}}{2} e^{-i2\omega_{p0}\tau_{ab}} \int d\omega_s \left[ a_{s,c}^\dagger(\omega_s) a_{i,d}^\dagger(2\omega_{p0}-\omega_s) e^{-i\omega_s\tau} + a_{s,d}^\dagger(\omega_s) a_{i,c}^\dagger(2\omega_{p0}-\omega_i) e^{-i(2\omega_{p0}-\omega_s)\tau} \right] |0\rangle, \quad (S13)$$

For this BS, the following unitary transformations are satisfied,

$$\begin{aligned}
a_{s,c}^\dagger(\omega_s) &\to [a_{s,e}^\dagger(\omega_s) + a_{s,f}^\dagger(\omega_s)]/\sqrt{2}, \\
a_{s,d}^\dagger(\omega_s) &\to [a_{s,e}^\dagger(\omega_s) - a_{s,f}^\dagger(\omega_s)]/\sqrt{2}, \\
a_{i,c}^\dagger(2\omega_{p0}-\omega_s) &\to [a_{i,e}^\dagger(2\omega_{p0}-\omega_s) + a_{i,f}^\dagger(2\omega_{p0}-\omega_s)]/\sqrt{2}, \\
a_{i,d}^\dagger(2\omega_{p0}-\omega_s) &\to [a_{i,e}^\dagger(2\omega_{p0}-\omega_s) - a_{i,f}^\dagger(2\omega_{p0}-\omega_s)]/\sqrt{2},
\end{aligned} \quad (S14)$$

where $a_{s,e}^\dagger(\omega_s)$ and $a_{s,f}^\dagger(\omega_s)$ ($a_{i,e}^\dagger(2\omega_{p0}-\omega_s)$ and $a_{i,f}^\dagger(2\omega_{p0}-\omega_s)$) are the creation operators of signal (idler) photons exiting from the output ports e and f of the BS, shownin Fig. S2b. With the unitary transformations in Eq. (S14), the $|\psi_4\rangle$ in Eq. (S13) evolves into,

$$|\psi_5\rangle = \frac{1}{\sqrt{2}}[|\psi_5^{ab}\rangle + |\psi_5^b\rangle],$$

$$|\psi_5^b\rangle = \frac{1}{2}\sqrt{\frac{\mu_c}{2}} e^{-i2\omega_{p0}\tau_{ab}} \int d\omega_s \left(e^{-i(2\omega_{p0}-\omega_s)\tau} + e^{-i\omega_s\tau}\right) [a_{s,e}^\dagger(\omega_s)a_{i,e}^\dagger(2\omega_{p0}-\omega_s) - a_{s,f}^\dagger(\omega_s)a_{i,f}^\dagger(2\omega_{p0}-\omega_i)], \quad (S15)$$

$$|\psi_5^{ab}\rangle = \frac{1}{2}\sqrt{\frac{\mu_c}{2}} e^{-i2\omega_{p0}\tau_{ab}} \int d\omega_s \left(e^{-i(2\omega_{p0}-\omega_s)\tau} - e^{-i\omega_s\tau}\right) [a_{s,e}^\dagger(\omega_s)a_{i,f}^\dagger(2\omega_{p0}-\omega_s) - a_{s,f}^\dagger(\omega_s)a_{i,e}^\dagger(2\omega_{p0}-\omega_i)].$$

As shown in Fig. S2b, a pair of DWDMs are used to selecte photons from the ports *e* and *f*, respectively. The photons exiting from the filters are detected by two SNSPDs after two polarization controllers, and the coincidence counts between the photons arriving at the two SNSPDs can be expressed as

$$n_{si,ef} = \eta_s \eta_i \int_T dt \int_{-\Delta\tau_{si}/2}^{\Delta\tau_{si}/2} d\tau_{si} |\langle\psi_5|a_e^\dagger(t)a_f^\dagger(t+\tau_{si})a_f(t+\tau_{si})a_e(t)|\psi_5\rangle|^2, \quad (S16)$$

where $a_e(t) = \int d\omega_e f_e(\omega_e) a_e(\omega_e) e^{-i\omega_e t}/\sqrt{2\pi}$ and $a_f(t) = \int d\omega_f f_f(\omega_f) a_f(\omega_f) e^{-i\omega_f t}/\sqrt{2\pi}$ are the annihilation operators of photons exiting from the ports *e* and *f*, respectively, with $a_e(\omega_e)$ and $a_f(\omega_f)$ being the corresponding presentations in frequency domain; $f_e(\omega_e)$ and $f_f(\omega_f)$ are the amplitude of the transmission function of the DWDMs after e and f, respectively. We can express them as $f_{e,f}(\omega_{e,f}) = \text{rect}((\omega_{e,f}-\omega_{s0,i0})/\Omega)$ according to the measurement of transmission spectra of the DWDMs after the ports *e* and *f*. The operators $a_e(\omega_e)$, $a_f(\omega_f)$, $a_{s,e}(\omega_s)$, $a_{i,e}(\omega_i)$, $a_{s,f}(\omega_s)$ and $a_{i,f}(\omega_i)$ satisfy the commutation relationships

$$\begin{aligned}[a_e(\omega_e), a_{s,e}(\omega_s)] &= \delta(\omega_e - \omega_s), \\ [a_e(\omega_e), a_{i,e}(\omega_i)] &= \delta(\omega_e - \omega_i), \\ [a_f(\omega_f), a_{s,f}(\omega_s)] &= \delta(\omega_f - \omega_s), \\ [a_f(\omega_f), a_{i,f}(\omega_i)] &= \delta(\omega_f - \omega_i).\end{aligned} \quad (S17)$$

Comparing to Eqs. (S16) and (S17), we can find that the component $|\psi_5^b\rangle$ in $|\psi_5\rangle$ has no contribution to the calculation of $n_{si,ef}$. Thus, Eq. (S16) evolves into

$$n_{si,ef} = \eta_s \eta_i \int_T dt \int_{-\Delta\tau_{si}/2}^{\Delta\tau_{si}/2} d\tau_{si} |\langle 0|a_f(t+\tau_{si})a_e(t)|\psi_5^b\rangle|^2. \quad (S18)$$

By substituting Eqs. (S15) and (S17) into Eq. (S18), we can finally obtain

$$n_{si,ef} = \frac{\eta_s \eta_i \mu_c}{8} [1 + \text{sinc}(\Omega\tau)\cos(\delta\omega_{si}\tau)], \quad (S19)$$

where $\delta\omega_{si} = 2(\omega_{p0} - \omega_{s0})$ is the spacing between the central angular frequencies of the two DWDMs in Fig. S2b. The Eq. (S19) gives a formula for us to fit the data of spatial quantum beating in Fig. 4a, except a visibility lower than 1 and a nonzero initial phase in the cosine function, resulting from the noise photons, the residual spatial path bunched photon-pairs, the imperfection of the BS used in Fig. S2b, and the instability of the delay $\tau$.

3. **Coincidence counts in the quantum state tomography of time-bin entanglement**

Table S1 Experimentally obtained coincidence counts

|     | Photon 1 | Photon 2 | $\|DD\rangle$ | $\|DR\rangle$ | $\|RD\rangle$ | $\|RR\rangle$ | $n_i$ |
| --- | --- | --- | --- | --- | --- | --- | --- |
| 1 | $\|1\rangle$ | $\|1\rangle$ | 3658 | 3743 | 3778 | 3632 | 14811 |
| 2 | $\|1\rangle$ | $\|2\rangle$ | 61 | 28 | 37 | 31 | 157 |
| 3 | $\|1\rangle$ | $\|D\rangle$ | 4365 | - | 4291 | - | 8656 |
| 4 | $\|1\rangle$ | $\|R\rangle$ | - | 4241 | - | 4339 | 8580 |
| 5 | $\|2\rangle$ | $\|1\rangle$ | 27 | 28 | 27 | 23 | 105 |
| 6 | $\|2\rangle$ | $\|2\rangle$ | 5217 | 5205 | 5225 | 5146 | 20793 |
| 7 | $\|2\rangle$ | $\|D\rangle$ | 4577 | - | 4410 | - | 8987 |
| 8 | $\|2\rangle$ | $\|R\rangle$ | - | 4562 | - | 4355 | 8917 |
| 9 | $\|D\rangle$ | $\|1\rangle$ | 5620 | 5696 | - | - | 11316 |
| 10 | $\|D\rangle$ | $\|2\rangle$ | 4582 | 4514 | - | - | 9096 |
| 11 | $\|D\rangle$ | $\|D\rangle$ | 16545 | - | - | - | 16545 |
| 12 | $\|D\rangle$ | $\|R\rangle$ | - | 5165 | - | - | 5165 |
| 13 | $\|R\rangle$ | $\|1\rangle$ | - | - | 5797 | 5663 | 11460 |
| 14 | $\|R\rangle$ | $\|2\rangle$ | - | - | 4629 | 4516 | 9145 |
| 15 | $\|R\rangle$ | $\|D\rangle$ | - | - | 4843 | - | 4843 |
| 16 | $\|R\rangle$ | $\|R\rangle$ | - | - | - | 1329 | 1329 |

We undertake four sets of projection measurements and the accumulation time of the coincidence counts is 10 s in all of the sets. The coincidence counts obtained in the four sets are summarized in Table S1. The 1st column defines the number of the projection basis, with the states in each row in the 2nd and 3rd columns being the states of the signal and idler photons in the corresponding basis, respectively. Totally, 16 bases are used and the values in each row in the 4th, 5th, 6th and 7th columns are the coincidence counts when the two-photon state is projected on the corresponding basis. The additional phase differences $\alpha$ and $\beta$ in the UMZIs in Fig. 2c are set at 0&0, 0&π/2, π/2&0, and π/2&π/2 in the 4th, 5th, 6th and 7th columns, respectively. A dash (-) indicates that coincidence count is absent in the corresponding projection measurement. The $n_i$ value in each row in the rightmost column is sum of the values in this row from 4th to 7th columns, and it is finally used to reconsctruc the density matrix of the photon pairs[6].

4. **Parameters of SNSPDs**

We use a set of SNSPD system with 6 channels for all measurements at single photon level. The SNSPD devices are manufactured in the Shanghai Institute of Microsystem and Information Technology (SIMIT). The entire detection system is developed by PHOTEC Corp., which including SNSPD devices, cryostat system and electronic control system. The SNSPDs operate at ~2.2 K in the cryostat system with a dark counting rate of ~150 Hz and a time jitter of ~100 ps. The dead time of all detectors are less than 30 ns and the efficiencies of the 6 channels at 1.5 $\mu$m band from 63% to 72% with an average of 67%. Each channel is characterized and the data is listed in Table S2.

Table S2 Main parameters of SNSPDs

|  | Dark count rate (Hz) | Detection efficiency (%) |
| --- | --- | --- |
| Channel 1 | 150 | 68 |
| Channel 2 | 200 | 63 |
| Channel 3 | 150 | 68 |
| Channel 4 | 150 | 65 |
| Channel 5 | 150 | 67 |
| Channel 6 | 100 | 72 |